# Teaching Digital Manufacturing Experimenting Blended-Learning Models By Combining MOOC And On-site Workshops In FabLabs

**Ella Hamonic**
Independent Researcher
Paris, France
hamonic.ella@gmail.com

**Anja Hopma**
IMT Institut Mines Telecom
Paris, France
anja.hopma@imt.fr

**Baptiste Gaultier**
IMT Atlantique
Rennes, France
baptiste.gaultier@imt-atlantique.fr

**Denis Moalic**
IMT Atlantique
Rennes, France
denis.moalic@imt-atlantique.fr

**Abstract**
**Teaching digital manufacturing at scale using MOOCs has opened opportunities for IMT, a network of French graduate engineering schools, to work closely with a community of learners and educators in physical spaces called Fab Labs. By setting up a cohort of life-long learning trainees taking the MOOC online and attending hands-on in-person workshops, IMT to experiment blended learning models and hybrid skills certification for project-based STEM courses.**

**Context**
In 2014, IMT started offering a MOOC dedicated to digital manufacturing on the French national FUN-MOOC [1] Open edX platform. In these early days of French MOOCs, the first session was a 12 weeks course aiming to give the general public a chance to discover the transformative potential of Fab Lab technologies. Fab Labs are community-based spaces and technologies that act as "technical prototyping platform (…) and also as a platform for learning and innovation" [2].

After an open call for contributions, the course team at IMT decided to co-design and to co-develop the entire course content with active members of the Fab Lab community. The MOOC project goal was to produce online education resources by bringing together several contributors (15) who had experimented new ways of learning in Fab Labs, and which had local impact in physical spaces and communities.

As a result of these collaborations and of the vitality of the MOOC's community of learners (104 733 enrollments, 10 718 FUN Honor Code certificates delivered over 3 years), several Fab Labs in France and in the French speaking world started organizing face-to-face workshops based on the MOOC, inviting learners to join the Fab Lab, take the MOOC and solve collectively the different challenges proposed in the MOOC (at least 15 spaces in 4 different countries reported in the MOOC discussions). A strength which partly explains the success of this course is certainly the nature of the Fab Lab community, which core value is knowledge sharing in physical spaces. Blended learning models with MOOCs in Fab Labs were therefore emerging organically.

**Course iterations and continuous content improvements**
To better fit with learner's needs, the original MOOC was rebundled in 2016 into three 4 weeks long MOOC to build a series focusing on different Fab Lab technologies (Arduino, 3D modeling/printing and Internet of Things - IoT). In 2017, a fourth course was added to the series providing an introduction to Fab Lab principles for "non-makers", inviting learners to discover and visit the closest Fab Lab near their home and to connect with this community.

**Bridging the gap between the physical and digital world: lab-based MOOCs**
Teaching object-based technologies online, such as Arduino, IoT and 3D printing, brings challenges in terms of learning design. Learning how to use hardware technologies implies using learning by doing as a fundamental principle for instructional design [3], transforming the MOOC into a lab environment. Doing so, the course team was following in the footsteps of another STEM team, from the University of Texas, who choose to teach embedded systems with hands-on methods using a MOOC and achieved very positive results [4].

Thus, each week of the IMT MOOC proposes a hands-on activity in which learners produce something (for example: design a 3D model, write an Arduino program, connect an object to an API) and then share evidence of their production in a peer-review process. If project-based learning offers meaningful opportunities of learning by showing quickly to the learner that she is able to create artefacts, it is also highly demanding in terms of student workload, materials [5] and personal organization. From the educator's point of view, project-based pedagogies are usually resource intensive in terms of tutoring, especially in the MOOC environment where participants are driven by self-motivation and might need quick individual technical support to avoid drop out. When learners are facing problems, the Fab Lab workshops could fill this gap and act as a place where they can find material, resources, help and further documentation. As a result, Fab Labs spaces and community could become mentoring spaces to support achievement and learning persistence in MOOCs.

**Blending the MOOC series in Fab Labs**

In 2017, the IMT course team therefore decided to experiment with blended learning, more specifically the "flipped-classroom" model [6], whereby learners interact online with educational resources and then attend workshops for further face-to-face training.

In collaboration with the French national work agency and a network of 5 Fablabs in the Paris region, the IMT *FabNum hybrid training* was offered to 40 job seekers, mostly boomers executives in career change, keen to be introduced to the digital manufacturing revolution and innovative product management methods.

The experiment uses the series of four MOOCs on digital manufacturing, delivering 16 weeks of content, covering an equivalent of 45 hours of online material. Trainees had to complete the MOOC activities at home and then attend a half-day workshops for each "week" of MOOC. The face-to-face part was organized in Parisian Fab Labs and amounted to 13 in-person workshops.

Physical workshops were dedicated to answering questions and to solving more advanced activities than those proposed in the MOOC, taking advantage of experimenting with a wide range of machines and hardware components available on-site. Furthermore, trainees were expected to develop a personal project on top of the programme, and to pitch it to a jury by the end of the training session.

**Hybrid IMT certification and possible business model**

For the first cohort, at the end of the hybrid training, after in-person assessments by IMT instructors, 90% of learners obtained an IMT online secure certification, certifying both skill tracks (MOOC + on-site learning). For the first experiment, the training costs were covered by public grants.

The *FabNum hybrid training* is currently being assessed by the French national commission for professional certification. Obtaining this label will make possible for the trainees to fund the training with their personal account for lifelong learning [8], opening possibilities for a sustainable business model to the project, in a context where Fab Labs organizations are also looking for their own business model and sustainability.

**Next Steps: Training Trainers**

We are currently asking ourselves how we can better support, at scale, the community of Fab Labs and educators using the MOOC ressources for on-site workshops and learners asking for more advanced skills certification. We are therefore deploying a blended learning experiment on a larger scale, with a **training of trainers** model. In 2018, we plan to certify a cohort of 100 trainees on-site, experimenting pathways for scalability. To be specific, we are training managers of Fab Labs and accompanying them in re-training cohorts of trainees in their Fab Labs, for the rest of the year. We've done this so far in Dakar (Senegal) and in Paris (France). This promising new experiment raises instructional design and skill certification issues that are being addressed by our institution.

**Conclusion**

By developing a MOOC project with a pre-existing network of physical spaces, online and on-site learning opportunities happened to naturally intertwine. The demand for on-site workshops and certification is increasing, bringing challenges to invent new forms of hybrid university credentialing where learning occurs online and in non-formal education spaces. Fab Labs are one example of such spaces.

**Acknowledgments**

This IMT project in collaboration with MCD/Makery was made possible with the support of Région Ile-de-France, Ville de Paris and Patrick and Lina Drahi Foundation.